\documentclass[11pt,aps,prd,tightenlines,nofootinbib,longbibliography,superscriptaddress,notitlepage,floatfix]{revtex4-1}
\usepackage{xspace}
\usepackage[english]{babel}
\usepackage{amsfonts}
\usepackage{amsmath, amsthm, amssymb, mathrsfs}
\usepackage{graphicx}
\usepackage[left=2cm, right=2cm, top=3cm, bottom=2cm]{geometry}
\usepackage[pdfencoding=auto,psdextra]{hyperref}
\usepackage{enumerate}
\usepackage[normalem]{ulem}
\usepackage{color}
\usepackage{soul}
\usepackage{float}
\usepackage[caption=false]{subfig}
\usepackage{subfiles}
\usepackage[makeroom]{cancel}
\usepackage{array}
\usepackage{tabularx}

\begin{document}
 \title{Perspectives on the dynamics in a loop quantum gravity effective description of black hole interiors}

\author{Mehdi Assanioussi}
  \email{mehdi.assanioussi@desy.de}
  \affiliation{II. Institute for Theoretical Physics, University of Hamburg,\\ Luruper Chaussee 149, 22761 Hamburg, Germany.}

\author{Andrea Dapor}
  \email{adapor1@lsu.edu}
  \affiliation{Department of Physics and Astronomy, Louisiana State University, Baton Rouge, LA 70803, USA.}

\author{Klaus Liegener}
  \email{liegener1@lsu.edu}
  \affiliation{Department of Physics and Astronomy, Louisiana State University, Baton Rouge, LA 70803, USA.}

\begin{abstract}
\noindent In the loop quantum gravity context, there have been numerous proposals to quantize the reduced phase space of a black hole, and develop a classical effective description for its interior which eventually resolves the singularity. However, little progress has been made towards understanding the relation between such quantum/effective minisuperspace models and what would be the spherically symmetric sector of loop quantum gravity. In particular, it is not clear whether one can extract the phenomenological predictions obtained in minisuperspace models, such as the singularity resolution and the spacetime continuation beyond the singularity, based on results in full loop quantum gravity.
In this paper, we present an attempt in this direction in the context of Kantowski-Sachs spacetime, through the proposal of two new effective Hamiltonians for the reduced classical model. The first is derived using Thiemann classical identities for the regularized expressions, while the second is obtained as a first approximation of the expectation value of a Hamiltonian operator in loop quantum gravity in a semi-classical state peaked on the Kantowski-Sachs initial data. We then proceed with a detailed analysis of the dynamics they generate and compare them with the Hamiltonian derived in General Relativity and the common effective Hamiltonian proposed in earlier literature.
\end{abstract}

\maketitle

\section{Introduction}\label{sec1}
It is said that General Relativity (GR) predicts its own inconsistency in the form of singularities: these are unavoidable features of several GR solutions, and are expected to be cured by a more complete and fundamental theory of gravity -- perhaps a quantum theory. In the case of Loop Quantum Gravity (LQG) \cite{LQG1,LQG2,LQG3}, the best results we have are based on models in Loop Quantum Cosmology (LQC) \cite{LQC1,LQC2,LQC3} -- a LQG-inspired quantization of cosmological minisuperspaces -- in which the big bang singularity is replaced by a ``big bounce'' bridging a contracting classical universe with an expanding one via a region of high (but finite) curvature in which gravity becomes effectively repulsive \cite{APS1,APS2,APS3}.

The LQC program has also been directed towards the study of black hole singularities, due to the fact that the interior of a spherically symmetric black hole can be described in terms of a Kantowski-Sachs cosmological model \cite{KS66}. Using this observation, several models have been proposed using Ashtekar variables for the black hole interior \cite{LQC-BH0,LQC-BH0b,LQC-BH01,LQC-BH02,LQC-BH03,LQC-BH04,LQC-BH1,LQC-BH2,LQC-BH3,LQC-BH4,LQC-BH5,LQC-BH6,LQC-BH7,LQC-BH8,LQC-BH9,LQC-BH10,LQC-BH11,LQC-BH12,LQC-BH13,LQC-BH14,LQC-BH15,LQC-BH16,LQC-BH17,LQC-BH18,LQC-BH19,LQC-BH20,LQC-BH21,LQC-BH22,LQC-BH23,LQC-BH24,LQC-BH25,LQC-BH26}, and almost all of them find the same qualitative conclusion: the singularity is resolved, being replaced by a space-like transition hypersurface, to the past of which there is a trapped region (the black hole region) and to the future of which there is an anti-trapped region (the white hole region). In other words, the singularity is replaced by a black hole to ``white hole'' transition. There is also an alternative approach to construct loop effective black hole models \cite{BBY, BLLN}, which is based on the polymerization of a more general class of inhomogeneous solutions, then reducing to the homogeneous case while preserving the covariance, namely the anomaly freedom of the effective constraints algebra. In the models based on this alternative approach, a different phenomenology of the interior of the black hole arises, illustrated in the appearance of an inner horizon (similarly to the classical Reissner-Nordstr\"{o}m black hole) and the occurrence of a spacetime signature change, revealing an Euclidean spacetime region inside the black hole \cite{BBY, BLLN}.

Inspired from the results in standard LQC for a homogeneous and isotropic spacetime, the models from  \cite{LQC-BH5,LQC-BH8,LQC-BH20,LQC-BH21,LQC-BH22,LQC-BH26}, although differing in the details such as the choices of regularisation parameters $\epsilon$, all postulate an effective dynamics for the classical symmetry reduced system, obtained via replacing the reduced Ashtekar-Barbero  connection in GR Hamiltonian by a regularized expression $a\to \sin(\epsilon a)/\epsilon$.
However, there was another proposal to construct a quantum Hamiltonian for the homogeneous and isotropic LQC model which relies on the construction of the Hamiltonian operator in LQG \cite{ThieID, ThieID2, AAL,AALM,ALM}. Though considered in the early days of LQC \cite{Boj1, Boj2, Boj3, Boj4, LQC-BH04}, then proposed in the context of spherically symmetric quantum model \cite{LQC-BH03}, and later on in the context of effective LQC models \cite{Yongge}, the idea of mimicking the construction of the Hamiltonian operator in LQG for LQC models was put aside. This was due to the impression that it does not lead to any significant difference with respect to the symmetry reduction method applied in standard LQC models. Yet, this impression turned out to not be entirely correct. Indeed, recent works \cite{ADLPshort,ADLPlong} have shown that the standard effective Hamiltonian in the homogeneous and isotropic LQC model, and the effective Hamiltonian obtained mimicking LQG regularization, present a relative agreement with the standard ones in the regime of GR, but they also display significant differences, in particular for the pre-bounce branch of the universe. The question therefore arises: what is the situation in the black hole context and what are the differences which arise, if at all, with respect to the standard treatment in LQC effective black hole models? The current work is dedicated to analyzing this question. In particular, we shall consider four Hamiltonians, and compare the phase space dynamics they generate for the black hole interior (taking initial conditions at the black hole horizon):
i) GR reduced Hamiltonian, $H_{\rm cl}$, which gives rise to the singular Schwarzschild solution;
ii) the Hamiltonian studied in \cite{LQC-BH6}, $H^{(1)}_{\rm eff}$, obtained via the aforementioned replacement in the GR reduced Hamiltonian;
iii) a new proposal $H^{(2)}_{\rm eff}$ obtained using Thiemann identities;
iv) and the Hamiltonian $H^{(3)}_{\rm eff}$ obtained from the expectation value of a LQG Hamiltonian \cite{ThieID,ThieID2} in a coherent state peaked on Kantowski-Sacks initial data.

Before we proceed, some observations are in order: first, while in isotropic and homogeneous cosmology the proposals ($iii$) and ($iv$) coincide \cite{ADLPshort,ADLPlong} when using a cubic graph for the coherent state, there is no guarantee that this remains true in general (in fact, we shall see that it is not the case for Kantowski-Sachs spacetime). Hence, we treat them as two separate proposals here. Second, in all our investigations, we shall adopt the so-called $\mu_o$-scheme for the choice of regulators. While this is known to lead to quantum geometry effects at low curvature in all models considered so far, it remains the simplest testing ground for conceptual ideas. Our purpose is to identify common qualitative features (such as singularity resolution and black hole to ``white hole" transition) that are expected to survive after a more ``physical'' $\bar\mu$-scheme is adopted. We also point out that the only known choice within the full theory of LQG is that of $\mu_o$-scheme \cite{DLP19}.

The structure of the paper is the following. In Section \ref{sec2}, we start by briefly reviewing the Hamiltonian formulation of Kantowski-Sachs metrics in terms of Ashtekar-Barbero variables, identifying the canonical variables of this system and obtaining the form of the classical Hamiltonian $H_{\rm cl}$. This Hamiltonian is then regularized following the approach of standard LQC models, thereby obtaining the form of $H_{\rm eff}^{(1)}$. We finally follow the alternative construction based on \cite{DL,ADLPshort,ADLPlong} to find the new proposal $H_{\rm eff}^{(2)}$, which takes into account Thiemann identities at the minisuperspace level \cite{ThieID,ThieID2}. Section \ref{sec3} is dedicated to the derivation of $H_{\rm eff}^{(3)}$, that is, the effective Hamiltonian obtained from full LQG. First, we present the choice of graph, whose parameters are the inverse of the number of nodes in each of the three spatial directions, $\mu_{|i|}$ (with $i=1,2,3$). Then, we observe that the leading order in the semiclassical expansion of the expectation value of the LQG Hamiltonian coincides with the GR Hamiltonian as regularized on the graph, $H^\mu_{\rm eff}$. Subsequently, we identify the fundamental variables (i.e., holonomies and fluxes) that describe the discrete Kantowski-Sachs geometry, and finally approximately evaluate $H^\mu_{\rm eff}$ in this case. 
In Section \ref{sec4} we solve numerically the dynamics of the four Hamiltonians under consideration, starting with initial conditions at the black hole horizon. We numerically solve the dynamics in all the cases and make a detailed comparison of the induced evolutions, finding that all the effective Hamiltonians produce a black hole to ``white hole" transition. We also evaluate the mass associated to the ``white hole'' horizon as a function of the initial black hole mass and compare the different cases. Section \ref{sec5} concludes the paper with a discussion of the results.
\section{Effective Kantowski-Sachs a la LQC}\label{sec2}
The interior of a spherically symmetric black hole is characterzied by the fact that Schwarzschild radial coordinate $r$ becomes timelike. The topology of the spatial slices (i.e., the surfaces of constant $r$) in the interior is therefore $[0,1] \times \mathbb S^2$, and the line element reads (in natural units, $G = 1 = c$)
\begin{align} \label{Schwarz}
ds^2 = - \left(\dfrac{2M_{\rm bh}}{r} - 1\right)^{-1} dr^2 + \left(\dfrac{2M_{\rm bh}}{r} - 1\right) dt^2 + r^2 d\Omega^2
\end{align}
By changing coordinates $(r,t) \to (T,R)$, with $R = t/L_R$ and $T = T(r)$ ($L_R$ is a constant with dimension of length), we can express the metric in the form
\begin{align} \label{KSmetric}
ds^2 = -N(T)^2 dT^2 + L_R^2 f(T)^2 dR^2 + L_S^2 g(T)^2 d\Omega^2
\end{align}
where $f$, $g$ and the coordinates are dimensionless (the latter ranging in $R \in [0,1]$, $\theta \in [0,\pi]$ and $\varphi \in [0,2\pi]$ respectively), $N$ is the lapse function and $L_S$ is a constant with dimension of length. Comparison with (\ref{Schwarz}) reveals that
\begin{align} \label{class-conditions}
f(T)^2 = \dfrac{2M_{\rm bh}}{r} - 1, \ \ \ g(T)^2 = \frac{r^2}{L_S^2}, \ \ \ N(T)^2 dT^2 = \left(\dfrac{2M_{\rm bh}}{r} - 1\right)^{-1} dr^2
\end{align}

Let us consider (\ref{KSmetric}) in its generality, that is, keeping $f$, $g$ and $N$ general functions of $T$. Then, we have what is called a Kantowski-Sachs model. This model can be described as a Hamiltonian system using the Ashtekar-Barbero formulation of GR \cite{Ash86,Ash88,Bar95}. In this formulation, the system is described via a phase space coordinatized by dynamical variables $(a,b,p_a,p_b)$ and a Hamiltonian constraint $H_{\rm cl}(a,b,p_a,p_b)$ generating their evolution with respect to $T$. The relation between the dynamical variables and the metric components are set as
\begin{align} \label{KSvar}
|p_a| = L_S^2 g^2, \ \ \ \ \ |p_b| = 2 L_R L_S |fg|
\end{align}
while the conjugated variables $a$ and $b$ satisfy the Poisson algebra
\begin{align}
\{a,p_a\} = \dfrac{\kappa \beta}{8\pi}, \ \ \ \ \ \{b,p_b\} = \dfrac{\kappa \beta}{8\pi}
\end{align}
where $\kappa = 16\pi G/c^3$. The explicit form of Ashtekar-Barbero variables is
\begin{align}\label{sympl_reduction}
\begin{array}{c}
A^1_1 = - a, \ \ \ A^2_2 = - b, \ \ \ A^3_3 = - b \sin\theta, \ \ \ A^1_3 = -\cos\theta
\\
\\
E^1_1 = |p_a| \sin\theta, \ \ \ E^2_2 = \dfrac{p_b}{2} \sin\theta, \ \ \ E^3_3 = \dfrac{p_b}{2}
\end{array}
\end{align}
where $\beta$ is the Barbero-Immirzi parameter of LQG. It follows that the Hamiltonian reads
\begin{align} \label{H-class}
H_{\rm cl} = -N\dfrac{8\pi}{\kappa \beta^2} \dfrac{{\rm sgn}(p_b)}{\sqrt{|p_a|}} \left[2 a b |p_a| + (b^2 + \beta^2) \dfrac{p_b}{2}\right]
\end{align}
The equations of motion generated by this $H_{\rm cl}$ can be solved analytically by for a clever choice of lapse function (namely, $N = \text{sgn}(p_b) \sqrt{|p_a|}$), and give
\begin{align} \label{class-analytical-sol}
\begin{array}{c}
a(T) = a^0 \cos\left(\dfrac{T-T_0}{2}\right)^{-4\ {\rm sgn}(p_a^0)}, \ \ \ \ \ p_a(T) = p_a^0 \cos\left(\dfrac{T-T_0}{2}\right)^{4\ {\rm sgn}(p_a^0)}
\\
\\
b(T) = -\beta \tan\left(\dfrac{T-T_0}{2}\right), \ \ \ \ \ p_b(T) = \dfrac{2}{\beta} a^0 p_a^0\sin\left(T-T_0\right)
\end{array}
\end{align}
where $a(T_0) = a^0$ and $p_a(T_0) = p_a^0$ are integration constants, while $b(T_0) = 0 = p_b(T_0)$ identify $T = T_0$ with the black hole horizon. Also, the singularity is identified by the condition $p_a=0$, occurring at the point $T=\pi + T_0$, and it translates into a diverging spatial curvature.

To give physical meaning to the integration constants, we plug (\ref{class-analytical-sol}) in the metric components of Kantowski-Sachs line element. Using the equations for $p_a$ in \eqref{KSvar} and \eqref{class-analytical-sol}, and imposing the second condition in (\ref{class-conditions}), we identify $r = \sqrt{|p_a^0|} \cos[(T-T_0)/2]^{2\ {\rm sgn}(p_a^0)}$. At this point, note that since $r$ decreases as we go further inside the black hole, the sign of $p_a^0$ must be positive and therefore ${\rm sgn}(p_a^0)=+1$. It follows that
\begin{align}
L_R^2 f^2 = \dfrac{p_b^2}{4|p_a|} = \dfrac{(a^0)^2 p_a^0}{\beta^2} 4 \sin\left(\dfrac{T-T_0}{2}\right)^2 \cos\left(\dfrac{T-T_0}{2}\right)^{-2} = 4 \dfrac{(a^0)^2 p_a^0}{\beta^2} \left(\dfrac{\sqrt{p_a^0}}{r} - 1\right)
\end{align}
The first condition in (\ref{class-conditions}) then implies $\sqrt{p_a^0} = 2M_{\rm bh}$ and $a^0 = \pm \beta L_R/(4 M_{\rm bh})$. In other words, if we want to obtain dynamically the line element in Schwarzchild coordinates (with the choice $R = t/L_R$), we must choose initial conditions at the black hole horizon\footnote
{
One can check that with these values the last condition in (\ref{class-conditions}) is also satisfied.
} as
\begin{align}
\begin{array}{lr}
a(T_0)=a^0:=\pm\dfrac{\beta L_R}{4 M_{\rm bh}}, \ \ \ \ \ b(T_0)=b^0:=0, \ \ \ \ \ p_a(T_0)=p_a^0:=4 M_{\rm bh}^2, \ \ \ \ \ p_b(T_0)=p_b^0:=0
\end{array}
\end{align}
We observe that the expression for $a^0$ depends on the length scale $L_R$, which means that $a^0$ is not physical and can be set arbitrarily. Different values of $a^0$ correspond to different rescalings of the coordinate $R$.
In the following, we shall choose $a^0$ to be a fixed constant, positive so as to ensure that the flow of $H_{\rm cl}$ is parametrized by an increasing $T$ (i.e.\ $T\geq T_0$), and independent of the mass parameter $M_{\rm bh}$.

In the context of LQC, several quantizations of this model have been proposed \cite{LQC-BH1,LQC-BH2,LQC-BH3,LQC-BH4,LQC-BH5,LQC-BH6,LQC-BH7,LQC-BH8}. It has been conjectured \cite{LQC-BH6} that the qualitative description of the semiclassical dynamics in these quantum models can be reproduced by classical effective models. These effective models are defined on the classical phase space with a dynamics generated by an effective Hamiltonian obtained by appropriately modifying the classical Hamiltonian \eqref{H-class}. The common choice of effective Hamiltonian is obtained by the replacements
\begin{align} \label{polymerization}
a \ \to \ \dfrac{\sin(\mu_aa)}{\mu_a}, \ \ \ \ \ b \ \to \ \dfrac{\sin(\mu_bb)}{\mu_b}
\end{align}
where $\mu_a$ and $\mu_b$ are phase space functions determined by the quantum model. The effective Hamiltonian therefore reads
\begin{align} \label{H-eff-old}
H_{\rm eff}^{(1)} = -N\dfrac{8\pi}{\kappa \beta^2} \dfrac{{\rm sgn}(p_b)}{\sqrt{|p_a|}} \left[2 \dfrac{\sin(\mu_aa)}{\mu_a} \dfrac{\sin(\mu_bb)}{\mu_b} |p_a| + \left(\dfrac{\sin(\mu_bb)^2}{\mu_b^2} + \beta^2\right) \dfrac{p_b}{2}\right]
\end{align}
It was shown in \cite{LQC-BH6} that in the case where $\mu_a$ and $\mu_b$ are chosen to depend on the initial conditions\footnote
{
This choice is referred to as the {\it generalized $\mu_o$-scheme} \cite{LQC-BH9}, as opposed to the standard $\mu_o$-scheme where the $\mu$ parameters are fixed positive numbers (which we use in this article), and the $\bar{\mu}$-scheme where the $\mu$ parameters are chosen to depend explicitly on the phase space variables.
}, 
the solutions to the equations of motion with $H_\text{eff}^{(1)}$ display a resolution of the singularity and a transition from a black hole interior to a white hole like interior with a second horizon at an instant $T_{\rm wh}$. We will display more details about this effective model in section \ref{sec4}.

An alternative effective Hamiltonian can be obtained following the approach which was advocated in \cite{Yongge,DL} in the context of LQC, and which relies on the regularization of the classical Hamiltonian used in the LQG approach. 
In the context of LQC \cite{ADLPshort,ADLPlong} the alternative regularization gave rise to a dynamics significantly different from the standard one. It is based on the classical relations referred to as Thiemann identities, which are used to define the Hamiltonian constraint in the full theory. This procedure starts by decomposing the GR Hamiltonian into ``Euclidean'' and ``Lorentzian'' parts:
\begin{align}
H = \int_\sigma {\rm d}x^3\;N(x) \;( H_E(x) + H_L(x) )
\end{align}
where
\begin{align} \label{HE-HL}
H_E = \frac{4}{\kappa^2\beta}F^J_{ab}\epsilon^{abc}\{V,A^J_c\}, \ \ \ \ \ H_L = -\frac{1+\beta^2}{\beta^7}\frac{16}{\kappa^4}\epsilon_{JMN}K^M_aK^N_b\epsilon^{abc}\{V,A^J_c\}
\end{align}
with $F^J_{ab}$  the curvature of the connection, $V$ the volume of the whole spatial manifold and the extrinsic curvature $K^I_a$ given by
\begin{align} \label{K-continuum}
K^I_a(y)=\frac{2}{\kappa\beta^3}\{A^I_a(y), \{\int_\sigma {\rm d}x^3 H_E(x),V\}\}
\end{align}
However, in general symplectic reduction and regularization do {\it not} commute, in particular when the considered phase space function involves Poisson brackets. Indeed, consider the symplectic reduction $\omega: (A,E)\mapsto (a,p_a,b,p_b)$ defined in (\ref{sympl_reduction}). Then, there exist (at least) three inequivalent regularizations of the continuous expression $\omega(K^I_a(y))$:
\begin{itemize}
\item One possibility is to regularize the object at the level of the full theory: introducing the regularization map $\iota_{A,E}$ based on (\ref{holonomies_fluxes}), the expression (\ref{K-continuum}) is regularized as
\begin{align} \label{K-reg-LQG}
\iota_{A,E}[K^I_x(y)] = -\frac{4}{\kappa\beta^3 {\mu_x}} \text{Tr}\left[\tau^I h(e_{y,x})^\dag \{h(e_{y,x}), \{\iota_{A,E}[\int_\sigma {\rm d}x^3 H_E(x)],\iota_{A,E}[V]\}\}\right]
\end{align}
{where $h(e_{y,x})$ is the holonomy along the edge $e$ with a boundary point $y$ and direction $x$,} and the reduction is performed at the end, obtaining $\omega \circ \iota_{A,E}[K^I_x(y)]$.
\item Another option is to perform the reduction first, and afterwards regularize the resulting expression at the reduced level, by a map $\iota_1$ based purely on regularization (\ref{polymerization}), which acts on reduced phase space functions $f(a,b,p_a,p_b)$ and gives $f(\sin(a\mu_a)/\mu_a,\sin(b\mu_b)/\mu_b,p_a,p_b)$. In this case, one first computes the Poisson brackets in (\ref{K-continuum}) at the continuum reduced level: from
\begin{align}
\omega(H_E) = \dfrac{8\pi}{\kappa} \dfrac{\text{sgn}(p_b)}{\sqrt{|p_a|}} \left[2|p_a| a b + \dfrac{p_b}{2} (b^2 - 1)\right]
\end{align}
and using $\omega(V) = 2\pi |p_b| \sqrt{|p_a|}$, it follows that
\begin{align} \label{H-V-in-H1}
\{\omega(H_E), \omega(V)\} =4\beta \pi( a |p_a| + b p_b)
\end{align}
At this point, computing the Poisson bracket with $\omega(A^I_x)$, gives for the non-vanishing components of $K^I_x$
\begin{align}
K^1_1 =- a/\beta, \ \ \ \ \ K^2_2= -b/\beta\ \ \ \ \, K^3_3 =- b \sin(\theta)/\beta
\end{align}
which are finally regularized to obtain
\begin{align} \label{old-K}
\iota_1[K^1_1] =-\frac{1}{\beta} \dfrac{\sin(\mu_a a)}{\mu_a}, \ \ \ \ \ \iota_1[K^2_2]= -\frac{1}{\beta}, \ \ \ \ \ 
\iota_1[K^3_3] =-\frac{\sin(\theta)}{\beta} \dfrac{\sin(\mu_b b)}{\mu_b}
\end{align}
\item The third option consists of a mixture of the previous two: we regularize the expression (\ref{K-continuum}) at the reduced level, but taking inspiration on the full theory regularization (\ref{K-reg-LQG}). Explicitly, this regularization map $\iota_2$ gives
\begin{align} \label{K-reg-LQCnew}
\iota_2[K^I_x(y)] = -\frac{4}{\kappa\beta^3 {\mu_x}} \text{Tr}\left[\tau^I \tilde h(e_{y,x})^\dag \{\tilde h(e_{y,x}), \{\int_\sigma {\rm d}x^3 \iota_1[\omega(H_E(x))],\iota_1[\omega(V)]\}\}\right]
\end{align}
where $\tilde h(e_{y,x})$ are chosen as\footnote
{
Note that these holonomies $\tilde h$ correspond to the full theory holonomies evaluated at $\theta = \pi/2$ (see Section 3).
}
\begin{align}
\tilde h(e_{y,1}) = e^{-a\tau_1 \mu_a}, \ \ \ \ \ \tilde h(e_{y,2}) = e^{-b\tau_2 \mu_b}, \ \ \ \ \ \tilde h(e_{y,3}) = e^{-b\tau_3 \mu_b}
\end{align}
with $\tau^I = -i \sigma^I/2$ the generators of the Lie algebra of $SU(2)$, and ${\mu_1=\mu_a\ ,\ \mu_2=\mu_3=\mu_b}$. Following this scheme, instead of (\ref{H-V-in-H1}) we have
\begin{align} \label{H-V-in-H2}
\{\iota_1[\omega(H_E)], \iota_1[\omega(V)]\} =4\beta\pi(|p_a| \dfrac{\sin(\mu_aa)}{\mu_a} \cos(\mu_bb) + p_b \dfrac{\sin(\mu_bb)}{\mu_b} \dfrac{\cos(a\mu_a) + \cos(b\mu_b)}{2})
\end{align}
which, plugged into (\ref{K-reg-LQCnew}) and upon carrying out the last Poisson bracket, gives finally
\begin{align} \label{new-K}
\iota_2[K^1_1] & = {-\frac{1}{\beta} \dfrac{\sin(\mu_aa)}{\mu_a} \cos(\mu_bb),} \\ \iota_2[K^2_2] & = \iota_2[K^3_3] = {-\frac{1}{\beta} \dfrac{\sin(\mu_bb)}{\mu_b} \dfrac{\cos(a\mu_a) + \cos(b\mu_b)}{2}}
\end{align}
\end{itemize}
It is clear that these three expressions are different as long as the regulators are kept finite:
\begin{align}
\iota_1[K^I_x(y)] \neq \iota_2[K^I_x(y)]  \neq \omega\circ\iota_{A,E}[K^I_x(y)] 
\end{align}
This in turn implies that the corresponding regularized Hamiltonians will be different. As it can be checked by direct computation, $H^{(1)}_{\rm eff}$ is obtained using the map $\iota_1[K_x^I]$ in (\ref{HE-HL}). Similarly, from $\iota_2[K_x^I]$ one finds
\begin{align} \label{H-eff-new}
H_{\rm eff}^{(2)} & =N \frac{8\pi {\rm sgn}(p_b)}{\kappa \sqrt{|p_a|}} \left[2 \dfrac{\sin(\mu_aa)}{\mu_a} \dfrac{\sin(\mu_bb)}{\mu_b} |p_a| + \left(\dfrac{\sin(\mu_bb)^2}{\mu_b^2} - 1\right) \dfrac{p_b}{2}\right.
\\
\notag 
& \left. -\frac{1+\beta ^2}{\beta^2} \dfrac{\sin(\mu_b b)}{\mu_b} (\cos (\mu_a a) + \cos (\mu_b b)) \left(|p_a| \dfrac{\sin (\mu_a a)}{\mu_a} \cos(\mu_b b) + \dfrac{p_b}{8} \dfrac{\sin (\mu_b b)}{\mu_b} (\cos (\mu_a a) + \cos (\mu_b b))\right)\right]
\end{align}
Both effective Hamiltonians (\ref{H-eff-old}) and (\ref{H-eff-new}) converge to $H_{\rm cl}$ in the limit $\mu_a,\mu_b \to 0$, ensuring the recovery of the continuum limit.

The use of map $\omega \circ \iota_{A,E}$ -- which gives rise to what we later call $H_{\rm eff}^{(3)}$ -- requires some more technology, and will be developed in the next Section. Here, we observe that this avenue to obtain a loop-effective model corresponds to an attempt to derive the effective Hamiltonian from the full quantum theory.
\section{Effective Kantowski-Sachs from LQG}\label{sec3}
Our approach in the case of the interior region of a black hole is the same as the one adopted for flat homogeneous and isotropic cosmology \cite{DL,ADLPshort}. Namely, the starting point is to consider a semi-classical coherent state on the Hilbert space of loop quantum gravity $\mathcal H$, peaked on the classical configuration of interest (for more details on coherent states in LQG, see \cite{TW1,TW2,TW3,STW4,dasgupta,LS,BT07,BT07b,ZT15,GCS}). Then, one computes the leading order, in a semi-classical expansion, of the expectation value of the LQG Hamiltonian operator originally proposed by Thiemann in its graph non-changing version \cite{ThieID,ThieID2,AQG1,AQG2} on the chosen semi-classical state. The obtained leading order is what one considers as the Hamiltonian in the effective theory\footnote{For example, one could use the complexifier coherent states from \cite{TW1,TW2,TW3}. This has been done in \cite{dasgupta}, however - in contrast with the discretization considered here - with a different choice of coordinates and discretized phase space functions, i.e. gauge covariant fluxes. For further implications on these different fluxes see \cite{ThiVII00,LS19a}.}. {Such methodology has been realized for instance in the case of homogeneous cosmology (within quantum reduced loop gravity (QRLG) \cite{EmaCOSMO} and LQG \cite{DL,DLlong}). In fact, in homogeneous isotropic LQC such expectation value coincides with the effective Hamiltonian obtained by the rule (\ref{polymerization}), thereby motivating the replacement method \cite{Taveras}.}

The calculation of the expectation value of the LQG Hamiltonian operator on a semi-classical state is in general involved. However, the calculation of the leading order of the expectation value is easier thanks to the properties of the semi-classical state. Indeed, the computations reduce to the replacement in the Hamiltonian of holonomy and flux operators by their classical, discretised expressions $\iota_{A,E}(A),\iota_{A,E}(E)$. The commutators are replaced by Poisson brackets on the discretised phase-space. As explained in the third bullet point of the previous Section, these Poisson brackets are then computed at the level of the full theory, and only afterwards we perform the symplectic reduction with respect to the discretised geometry on which the semi-classical state is peaked.

The first step is therefore to perform a discretization of the spacetime manifold using a choice of graph and its dual $2$-complex. Since we are interested in a spacetime of the Kantowski-Sachs type, we choose the graph $\Gamma$ to be adapted to the cylindrical coordinates in which the Kantowski-Sachs metric is expressed \eqref{KSmetric}. Such choice of graph simplifies considerably the calculations.
Thus, we are interested in a fixed graph $\Gamma$, which is chosen to be a compact subset of the cubic lattice $\mathbb{Z}^3$ embedded in $[0,1]\times\mathbb{S}^2$. We choose a graph $\Gamma$ which has a finite number of vertices equal to the product $N_1N_{2}N_{3}$ where $N_1,\ N_{2},\ N_{3}\in\mathbb{N}$ are the numbers of vertices in the directions $R$, $\theta$ and $\varphi$ respectively. 
Finally, the cubic lattice $\Gamma$ is oriented the following way: at each vertex $v$ there are six edges $e_{v,i}$ ($i=\pm 1,\pm 2,\pm 3$ such that $1,2,3$ correspond to the directions $R$, $\theta$ and $\varphi$ respectively, while $+$ is for outgoing edges and $-$ for incoming edges) starting at $v$ and going along the directions $i$ (with constant coordinates along the remaining directions). The coordinate lengths of these edges are $\mu_1 L_R$, $\mu_2 L_S$ and $\mu_3 L_S$, where we define
\begin{align} \label{Mus}
\mu_1 = \dfrac{1}{N_1}, \ \ \ \ \ \mu_2 = \dfrac{\pi}{N_2+1}, \ \ \ \ \ \mu_3 = \dfrac{2\pi}{N_3}
\end{align}
Therefore, the coordinates $(R_v,\theta_v,\varphi_v)=:v$ of a generic vertex $v$ in the graph take values in
\begin{align}
R_v\in \bigg\{\mu_1, 2\mu_1,..., 1\bigg\},\hspace{10pt}\theta_v\in\bigg\{\mu_2, 2\mu_2,..., \pi - \mu_2\bigg\},\hspace{10pt}\varphi_v\in\bigg\{\mu_3, 2\mu_3,..., 2\pi\bigg\}\ .
\end{align}

In the following, we present the discrete version of the Kantowski-Sachs system on the graph $\Gamma$ and the corresponding holonomies and fluxes.
	\subsection{Discretization of Kantowski-Sachs}\label{sec31}
The discrete model of Kantowski-Sachs solution is obtained by introducing the holonomy-flux algebra as discrete objects associated with the edges of the graph $\Gamma$, given by the semi-classical state we chose and described above, and surfaces of the dual $2$-complex. These surfaces are obtained in the following way: at each edge $e_{v,i}$ we consider the surface $S_{e_{v,i}}$ orthogonal to it at the midpoint and oriented in the direction $i$, such that it has sides of coordinate lengths $\mu_{|j|}$ and $\mu_{|k|}$ ($j,k\neq \pm i$) parallel to the edges of the graph $\Gamma$. Then, we define the holonomies $h$ and the fluxes $E_I$ as
\begin{align}\label{holonomies_fluxes}
h(e_{v,i}):=\mathcal{P}\exp\left(\int_0^1ds\; \dot e_{v,i}^a(s)A^I_a(e_{v,i}(s))\tau_I\right),\hspace{20pt}E_I(e_{v,i}):=\int_{S_{e_{v,i}}}dx^a\wedge dx^b\;\epsilon_{abc} E^c_I(\vec{x})
\end{align}
where $\dot{e}_{v,i}$ is the normalized tangent vector to the edge $e_{v,i}$ at $v$, and we choose the explicit basis $\tau_I=-i\sigma_I/2$ of $\mathfrak{su}(2)$, i.e. $[\tau_I,\tau_J]=\epsilon_{IJK}\tau_K$.

The leading order of the expectation value of Thiemann Hamiltonian operator, which we denote $H_\text{eff}^{\mu}$, takes the form
\begin{align}
H_\text{eff}^{\mu}:= N(H^\mu_E + H^\mu_L) \,
\end{align}
where
\begin{align}
H^\mu_E:=\frac{-4}{\kappa^2\beta}\sum_{v\in V(\Gamma)}\frac{1}{T_v}\sum_{i,j,k}\epsilon(e_{v,i},e_{v,j},e_{v,k})\;tr\left(
(h(\Box^v_{ij})-h(\Box^v_{ij})^\dagger)h(e_{v,k})^\dagger\{h(e_{v,k}),V^\mu \}\right)
\end{align}
and
\begin{align}
H^\mu_L:=&\frac{64(1+\beta^2)}{\kappa^4\beta^7}\sum_{v\in V(\Gamma)}\frac{1}{T_v}\sum_{i,j,k}\epsilon(e_{v,i},e_{v,j},e_{v,k}) tr\left(h(e_{v,i})^\dagger\{h(e_{v,i}),K\}h(e_{v,j})^\dagger\{h(e_{v,j}),K\}h(e_{v,k})^\dagger\{h(e_{v,k}),V^\mu \}\right)
\end{align}
with $T_v$ being an averaging factor which depends on the valence of the vertex $v$, and $V^\mu$ being the discrete volume given as
\begin{align}
V^\mu:=\sum_{v\in \text{V}(\Gamma)}\sqrt{\frac{1}{T_v}|Q_v|},\hspace{20pt}Q_v=\sum_{i,j,k}\epsilon(e_{v,i},e_{v,j},e_{v,k})\epsilon^{IJK}E_{I}(e_{v,i})E_{J}(e_{v,j})E_{K}(e_{v,k})
\end{align}
and $K:=\{V^\mu ,H^\mu_E\}$. The factor $\epsilon(e_{v,i},e_{v,j},e_{v,k})=sgn(\det(\dot{e}_{v,i},\dot{e}_{v,j},\dot{e}_{v,k}))$ and $\text{V}(\Gamma)$ is the set of all vertices of the graph. Note that the six-valent vertices have $T_v=48$, while the five-valent vertices with $R_v=\mu_1, 1$ or $\theta_v=\mu_2, \pi-\mu_2$ have $T_v=24$. The symbol $h(\Box^v_{ij})$ corresponds to the holonomy, along the oriented loop $\Box^v_{ij}$, defined as
\begin{align}
 h(\Box^v_{ij}):= h(e_{v+\mu_{|j|}\dot{e}_{v,j},-j}) h(e_{v+\mu_{|i|}\dot{e}_{v,i}+\mu_{|j|}\dot{e}_{v,j},-i}) h(e_{v+\mu_{|i|}\dot{e}_{v,i},j}) h(e_{v,i})
\end{align}

The final expressions for the quantities $H^\mu_E$ and $H^\mu_L$ in terms of the phase space variables $\{a,b,p_a,p_b\}$ are obtained by first performing the calculation of the Poisson brackets involved on the level of the general holonomy-flux algebra, then evaluating the holonomies and fluxes for the Kantowski-Sachs metric. These are
\begin{align}
\begin{array}{ll}
E_I(e_{v,\pm 1})=\pm \delta_{I1} 2|p_a|\sin(\theta_v)\sin(\dfrac{\mu_2}{2})\mu_3\ , \qquad & \qquad h(e_{v,\pm 1})=\exp(\mp a \tau_1\mu_1 )\ ,
\\
\\
E_I(e_{v,\pm 2})=\pm \delta_{I2} \dfrac{p_b}{2}\sin(\theta_v \pm \dfrac{\mu_2}{2})\mu_1\mu_3\ , \qquad & \qquad h(e_{v,\pm 2})=\exp(\mp b\tau_{2} \mu_{2})\ ,
\\
\\
E_I(e_{v,\pm 3})=\pm \delta_{I3} \dfrac{p_b}{2}\mu_2\mu_1\ , \qquad & \qquad h(e_{v,\pm 3})=\exp(\mp \mu_{3}(b \sin(\theta_v)\tau_{3}-\cos(\theta_v)\tau_1))
\end{array}
\end{align}
These computations produce an analytic expression for $H_\text{eff}^{\mu}$ which is too lengthy to fit in an article, due to the lengthy expression of $H^\mu_L$. For the sake of the argument, we display the expression of $H^\mu_E$:
\begin{align}
H^\mu_E & = {\rm sgn}(p_b) \dfrac{2\pi}{\kappa \sqrt{|p_a|}} \sum_{\theta = \mu_2}^{N_2\mu_2} \mu_2 \bigg[\sqrt{\dfrac{\tan(\mu_2/2)}{\mu_2/2}} 2|p_a| \sin(\theta) \dfrac{\sin(a\mu_1)}{\mu_1}
\\
& \times \bigg(b \dfrac{\sin(2\mu_3\chi(\theta))}{2\mu_3\chi(\theta)} + \cos(\dfrac{\mu_2}{2}) \dfrac{\sin( b \mu_2)}{\mu_2} \dfrac{b^2 \sin(\theta)^2 + \cos(\theta)^2\cos(\mu_3\chi(\theta))}{b^2 \sin(\theta)^2 + \cos(\theta)^2}\bigg) \notag
\\
& + \sqrt{\dfrac{\mu_2}{2} \cot(\dfrac{\mu_2}{2})} \dfrac{p_b}{2} \bigg(\cos(\theta) \dfrac{\sin(\mu_3\chi(\theta))}{\mu_3\chi(\theta)} \dfrac{\cos(\mu_3\chi(\theta-\mu_2)) - \cos(\mu_3\chi(\theta+\mu_2))}{\mu_2} \notag
\\
& + \cos(\mu_3\chi(\theta)) \dfrac{1}{\mu_2} \sum_{s = \pm 1} s\dfrac{\sin(\mu_3\chi(\theta + s\mu_2))}{\mu_3\chi(\theta + s\mu_2)} \big(\cos(\theta + s\mu_2)\cos(b \mu_2) + b \sin(\theta + s\mu_2)\sin(b \mu_2)\big)\bigg)\bigg] \notag
\end{align}
where $\chi(\theta) := \frac{1}{2} \sqrt{b^2 \sin(\theta)^2 + \cos(\theta)^2}$. Unfortunately, the complexity of the expression of $H_\text{eff}^{\mu}$ makes it at the moment impossible to solve the equations of motion neither analytically nor numerically. However, since conceptually we are interested in large graphs, the values of the parameters $\mu_{|i|}$ are considered to be very small compared to unity (see \eqref{Mus}). Hence, we take the power series expansion of $H_{\rm eff}^{\mu}$ in $\mu_{|i|}$. 
In particular, we evaluate the expansion up to second order in $\mu_{|i|}$, giving the following expression
\begin{align}
H_{\rm eff}^{(3)} & := 
-N \dfrac{8\pi\ {\rm sgn}(p_b)}{\kappa \beta^2 \sqrt{|p_a|}} \bigg[2 a b |p_a| + (b^2 + \beta^2) \dfrac{p_b}{2}\bigg]
\\
& + N \dfrac{\pi\ {\rm sgn}(p_b)}{144 \kappa \beta^2 \sqrt{|p_a|}}  \bigg[96 \mu_1^2 a^2 b \bigg(2 a \left(3 \beta ^2+5\right) |p_a|+3 b \left(\beta ^2+1\right) p_b \bigg) \notag
\\
& + 24 \mu_2^2 \bigg(2 a b |p_a| \left(18 b^2 \beta ^2+2 \left(11 b^2+6 \beta ^2\right) +17 \right)+p_b \left(2 \left(3 \beta ^2+5\right) b^4+3 b^2+7\beta^2\right)\bigg) \notag
\\
& + \mu_3^2 \bigg(2 a b |p_a| \left(288 b^2 \beta ^2+\left(352 b^2+3\beta^2\right) +59 \right)+b^2 p_b \left(96 b^2 \beta ^2+5 \left(32 b^2+9 \beta^2\right) -19 \right) \bigg)\bigg] \notag
\end{align}
\section{Comparative Analysis of the Effective Dynamics}\label{sec4}
In this section we expose the analysis of the dynamics in the effective Kantowski-Sachs models described by the three Hamiltonians $H_\text{eff}^{(1)}$, $H_\text{eff}^{(2)}$ and $H_\text{eff}^{(3)}$ presented in the previous sections. Specifically, we compare the solutions to the equations of motion induced by the aforementioned Hamiltonians and we discuss the qualitative aspects featuring in each case: in particular, the resolution of the singularity and the ``white hole''-like region.

The equations of motion that we are solving for the variables $a$, $b$, $p_a$ and $p_b$ are of the form
\begin{align}\label{EqM}
 \dot a(T) = \{a(T),H_\text{eff}^{(i)}\}\ ,\ \dot b(T) = \{b(T),H_\text{eff}^{(i)}\}\ ,\ \dot p_a(T) = \{p_a(T),H_\text{eff}^{(i)}\}\ ,\ \dot p_b(T) = \{p_b(T),H_\text{eff}^{(i)}\}\ ,
\end{align}
where $i$ ranges from $1$ to $3$. With a suitable choice of lapse function, these equations can be solved analytically in the case of the Hamiltonian $H_\text{eff}^{(1)}$ \cite{LQC-BH1, LQC-BH6}, but this is so far not the case for the other two Hamiltonians. However, one can proceed with solving the equations of motion numerically with a lapse function fixed as earlier $N=\text{sgn}(p_b) \sqrt{|p_a|}$ and with initial conditions taken as discussed in section \ref{sec2}, namely
\begin{align}\label{InitCond}
\begin{array}{lr}
a^0=const., \ \ \ \ \ b^0=0, \ \ \ \ \ p_a^0=4 M_{\rm bh}^2, \ \ \ \ \ p_b^0=0
\end{array}
\end{align}
then compare the solutions for the various dynamics considered.
As already discussed, the initial conditions \eqref{InitCond} are set to reproduce Schwarzschild solution at the horizon, allowing us to interpret the resulting dynamics as the (effective) evolution of the interior of a Schwarzschild black hole. It is important to remember that the value of $a^0$ is in principle free, and we will discuss the relevance of the choice of $a^0$ in the effective models later in this section.

\begin{figure}[!h]
	\begin{centering}
		\subfloat[]{\includegraphics[width=0.8\textwidth]{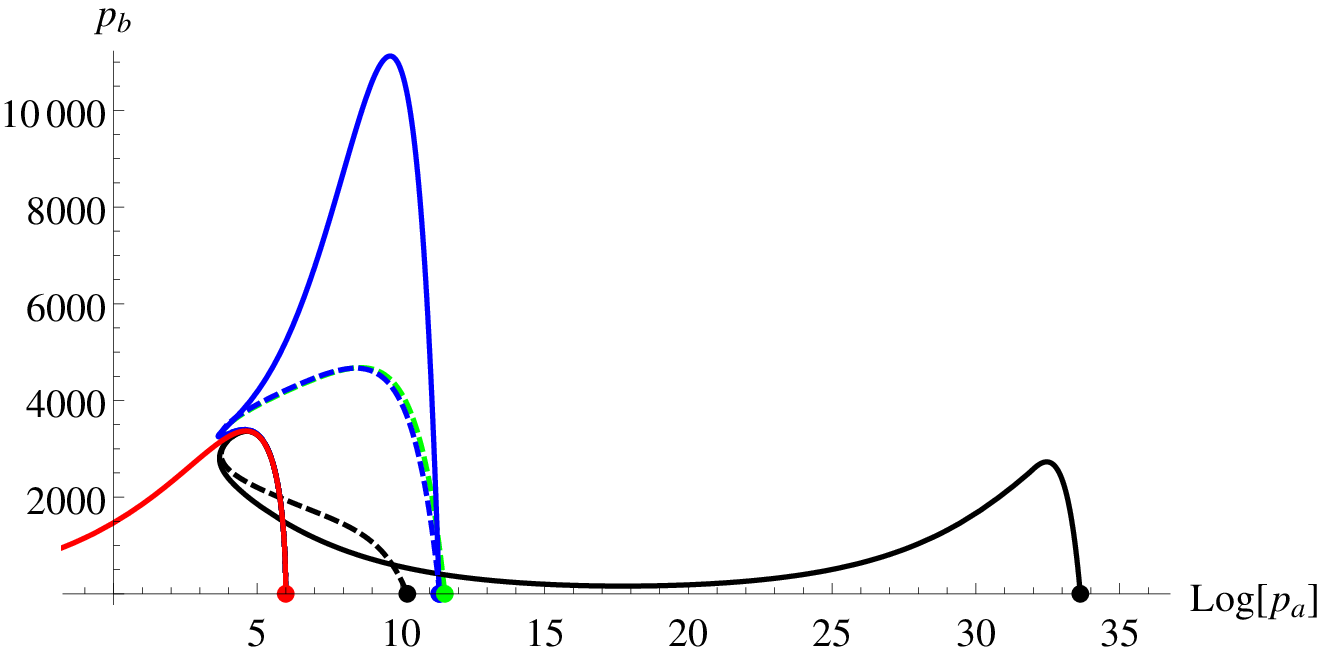}}
		\hspace{0.3cm}
		\subfloat[]{\includegraphics[width=0.8\textwidth]{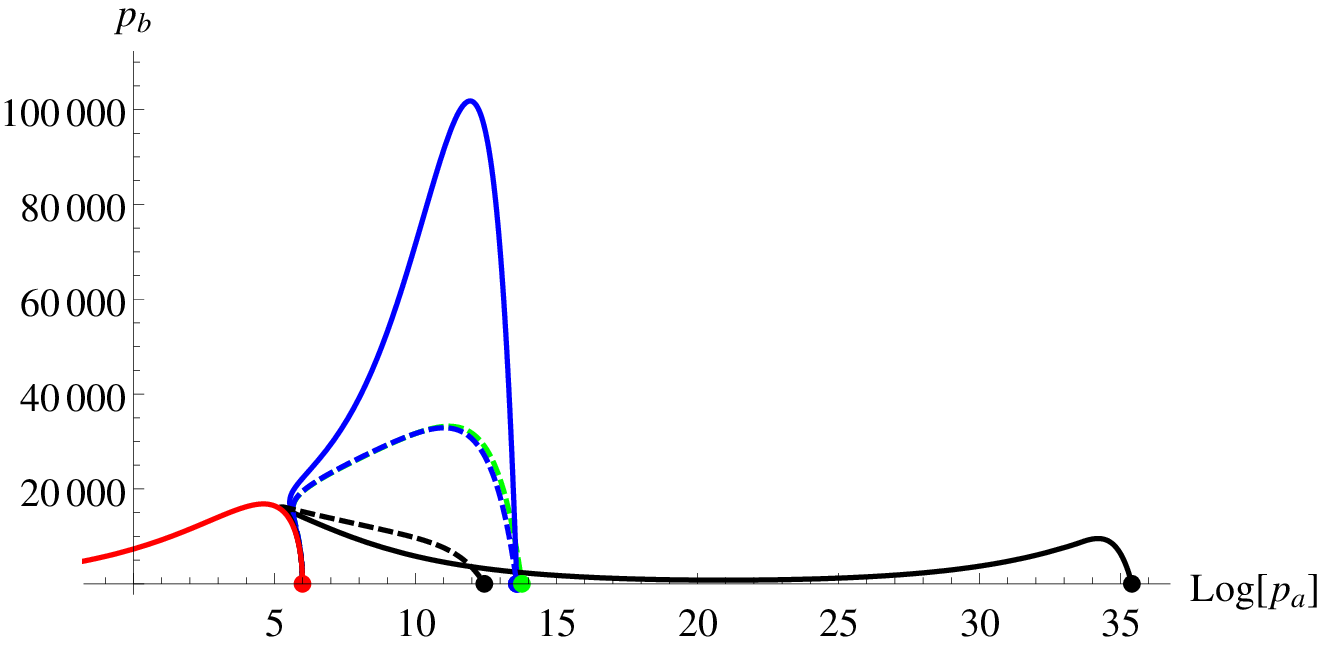}}
	\end{centering}
	\caption{A parametric plot displaying the evolution of $p_b(T)$ as a function of $\log[p_a(T)]$ set at the instant $T_0=0$ ($\log[p_a(T_0)]\approx 6,\ p_b(T_0)=0$, the red dot), with $M_{\rm bh}=10$ in Planck units and in (a) $a^0=1$ while in (b) $a^0=5$. The continuous red, black and blue trajectories correspond respectively to the solutions obtained using the Hamiltonians $H_{\rm cl}$, $H_\text{eff}^{(1)}$, $H_\text{eff}^{(2)}$. The dashed green trajectory corresponds to the solutions obtained using $H_\text{eff}^{(3)}$, while the dashed black and blue trajectories correspond to the solutions obtained using Hamiltonians defined as the second order expansions in $\mu$'s of $H_\text{eff}^{(1)}$ and $H_\text{eff}^{(2)}$ respectively. The remaining dots correspond to the white hole horizons obtained for each trajectory. We take $G=c=1$, $\beta=0.2375$, $\mu_a = \mu_b = 1/10$.}
	\label{fig1}
\end{figure}

The other important element of our analysis is the choice of $\mu_a$ and $\mu_b$. While it might be argued that a $\bar\mu$-scheme (in which these regulators are phase space functions) is more physical, there is currently no agreement on the correct choice. Moreover, as pointed out in \cite{LQC-BH9}, the choice of $\mu$'s as dependent on the initial conditions (that is the generalized $\mu_0$-scheme, for example $\mu_b = \sqrt{\Delta/p_a^0}$ adopted in \cite{LQC-BH6}), implicitly assumes that $\mu$'s are constants of motion; however, the fact that the regulator is a phase space function (albeit conserved by the dynamics) implies that the equations of motion for the fundamental variables $(a,b,p_a,p_b)$ are different from those computed in the $\mu_0$-scheme (i.e., with $\mu$'s constant on the whole phase space). While it might be possible to circumvent these issues by relying on an extended phase space (where the regulator become themselves new phase space coordinates), we prefer to refrain from introducing these complications, and thus stick to the original $\mu_0$-scheme. From the point of view of the full theory, this is so far the only choice, since in $H_{\rm eff}^{(3)}$ the regulators $\mu_1$, $\mu_2$ and $\mu_3$ are given in terms of the numbers of nodes \eqref{Mus}. For our analysis, we use the identification
\begin{align}\label{Choice_Mus}
 \mu_1 = \mu_a = \mu_a^o , \qquad \mu_2 = \mu_3 = \mu_b= \mu_b^o,
\end{align}
where $\mu_a^o$ and $\mu_b^o$ are fixed positive numbers. Now, we can proceed with the analysis of the effective evolution. In figure \ref{fig1} we display the evolution trajectories of $p_b$ as a function of $\log(p_a)$, obtained for an initial mass $M_{\rm bh}=10$ in Planck units and with the initial conditions \eqref{InitCond} set at the instant $T_0=0$ where all effective and GR trajectories meet. To make the comparison with $H_{\rm eff}^{(3)}$ (which is an expansion to second order in $\mu$'s) more meaningful, we considered also the trajectories generated by the second-order expansions of $H_{\rm eff}^{(1)}$ and $H_{\rm eff}^{(2)}$.
The plots in figure \ref{fig1} show that all the effective solutions display a non-vanishing minimum for $p_a$ (reached at different times which we denote $T_{b}^{(i)}$), implying an absence of the singularity predicted by classical GR (red curve) and the occurrence of a bounce at the instants $T_{b}^{(i)}$. In particular, the trajectories of the expansion of $H_{\rm eff}^{(2)}$ and $H_{\rm eff}^{(3)}$ (dashed blue and dashed green, respectively) agree very well, indicating that -- at least to second order in $\mu$'s -- the two proposals essentially agree.
We also observe that all the effective trajectories reach a vanishing value for $p_b$ (also at different times $T_{\rm wh}^{(i)}$, other than $T_o$), indicating the presence of a Killing horizon in the effective solutions. In this, it is interesting to compare the trajectory of $H_{\rm eff}^{(2)}$ and its expansion (blue lines, solid and dashed respectively): while the two curves are very different, they end up in the same final value for $\log(p_a)$ (which is also close to the final value obtained from $H_{\rm eff}^{(3)}$. The same is not true for $H_{\rm eff}^{(1)}$ and its expansion (black lines), whose trajectories are extremely different and end up in different final values.
Through the evaluation of the geodesics expansions\footnote
{
Expansion parameters are defined as $\Theta_\pm = h^{\alpha\beta} \nabla_\alpha k_\beta^{\pm}$ where $h_{\alpha\beta}$ is the metric induced by $g_{\mu\nu}$ on the 2-spheres coordinatized by $(\theta,\varphi)$ and $k^\pm$ are the vector fields tangent to the congruences of outgoing and ingoing radial null geodesics -- given here by $k_a^+ = 1/\sqrt{2}(-N,L_R f,0,0)$ and $k_a^- =1/\sqrt{2} (-N,-L_R f,0,0)$ respectively.
}
$\Theta_{\pm}=\dot p_a/(\sqrt{2}p_a N)$ associated to the two future-oriented null normals to the surfaces of $T$ and $R$ constant, we observe that $\Theta_{\pm}$ have the same signs throughout the evolution and correspond to the sign of $\dot p_a /N$ (since $p_a$ remains positive throughout the evolution), which is independent of the choice of the coordinates. Namely, it is negative in the region $0<T<T_{b}$ and positive in the region $T_{b}<T<T_{\rm wh}$, while vanishing at $T_{b}$. Therefore the boundary $T = T_{b}$ is a transition surface from a trapped region -- the black hole interior -- to an anti-trapped region -- the ``white hole''-like interior.

\begin{figure}[h!]
\centering
\subfloat[]{\includegraphics[width=8.5cm]{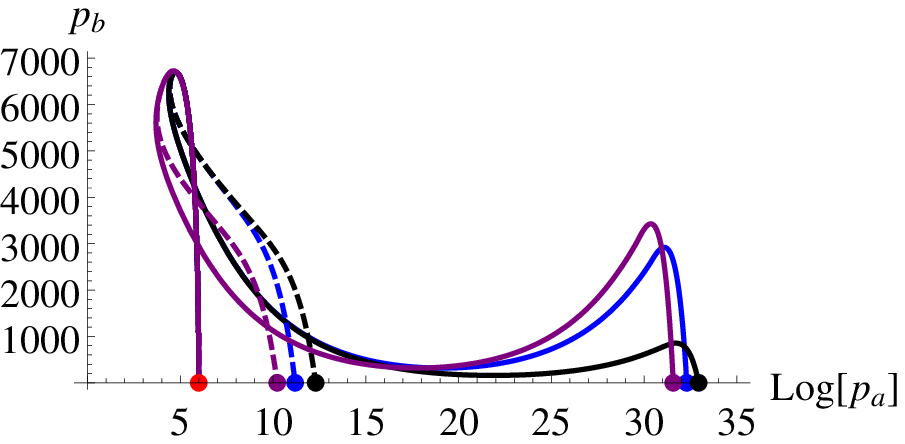}}
\hspace{0.3cm}
\subfloat[]{\includegraphics[width=8.5cm]{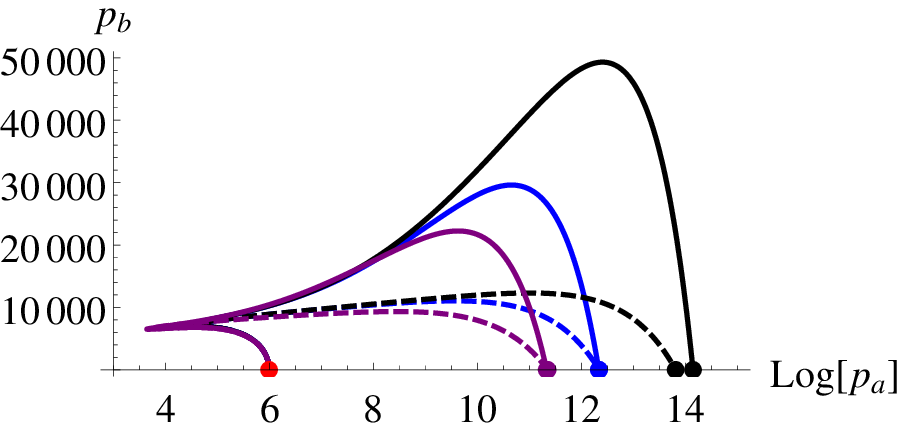}}
\vspace{0.3cm}
\subfloat[]{\includegraphics[width=8.5cm]{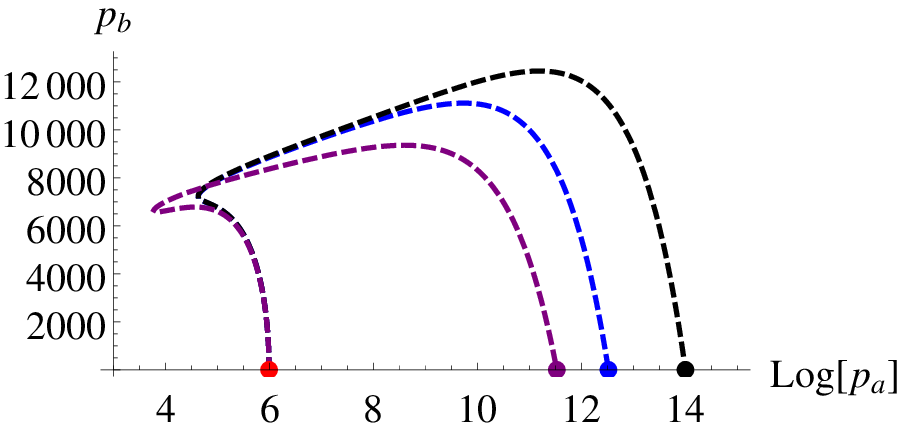}}
\caption{Parametric plots displaying the evolution of $p_b(T)$ as a function of $\log[p_a(T)]$, using the initial conditions \eqref{InitCond} set at the instant $T_0=0$ ($\log[p_a(T_0)]\approx 6,\ p_b(T_0)=0$, the red dot), with $a^0=2$ and $M_{\rm bh}=10$ in Planck units, evaluated for different values of $\mu_a$ and $\mu_b$: a) solutions obtained using the Hamiltonian $H_\text{eff}^{(1)}$ (continuous) and its expansion (dashed); b) solutions obtained using the Hamiltonian $H_\text{eff}^{(2)}$ (continuous) and its expansion (dashed); c) solutions obtained using the Hamiltonian $H_\text{eff}^{(3)}$ (dashed). 
The blue trajectories correspond to $\mu_a=1/10$ and $\mu_b=1/10$, the black trajectories correspond to $\mu_a=1/10$ and $\mu_b=1/20$ and the purple trajectories correspond to $\mu_a=1/20$ and $\mu_b=1/10$. We take $G=c=1$, $\beta=0.2375$.}
\label{fig2}
\end{figure}

In figure \ref{fig2} we display the trajectories $p_b$ vs $\log(p_a)$ for various values of $\mu_a$ and $\mu_b$ obtained from (a) $H^{(1)}_{\rm eff}$, (b) $H^{(2)}_{\rm eff}$ and (c) $H^{(3)}_{\rm eff}$. We observe that, while the overall shape of the trajectories of each effective model does not change dramatically as one changes the values of $\mu$'s, a common effect can be detected: the final point of the curve (which coincides with the white hole horizon) shifts to the left as $\mu_a$ is smaller (purple lines) and to the right as $\mu_b$ is smaller (black lines). Comparison of panels (b) and (c) also confirms the similarity between the trajectories of expanded $H_{\rm eff}^{(2)}$ and $H_{\rm eff}^{(3)}$, which remains true for different choices of $\mu$'s.

The presence of a bounce and a second horizon were already observed in the analysis of the dynamics generated by $H_\text{eff}^{(1)}$ in \cite{LQC-BH6} -- although we recall that here we are using a different choice of regulators. While the bounce seems to be a generic feature, the presence of a second Killing horizon is more subtle. For example, in the context of QRLG \cite{EmaBH, EmaDY} and within a specific $\bar \mu$-scheme, no second horizon was present in the solutions to the effective dynamics. 
In presence of a second horizon, an interesting feature to analyze is the mass $M_{\rm wh}$ associated to such horizon, and investigate its possible dependence on the initial mass of the black hole $M_{\rm bh}$. The white hole mass is given by
\begin{align}
M_{\rm wh}^{(i)} := \dfrac{\sqrt{p_a^{(i)}(T_{\rm wh}^{(i)})}}{2} 
\end{align}
where $T_{\rm wh}^{(i)}$ is the value of $T \neq T_{0}$ at which $p_b^{(i)}$ vanishes.

\begin{figure}[!h]
	\begin{centering}
		\includegraphics[width=0.7\textwidth]{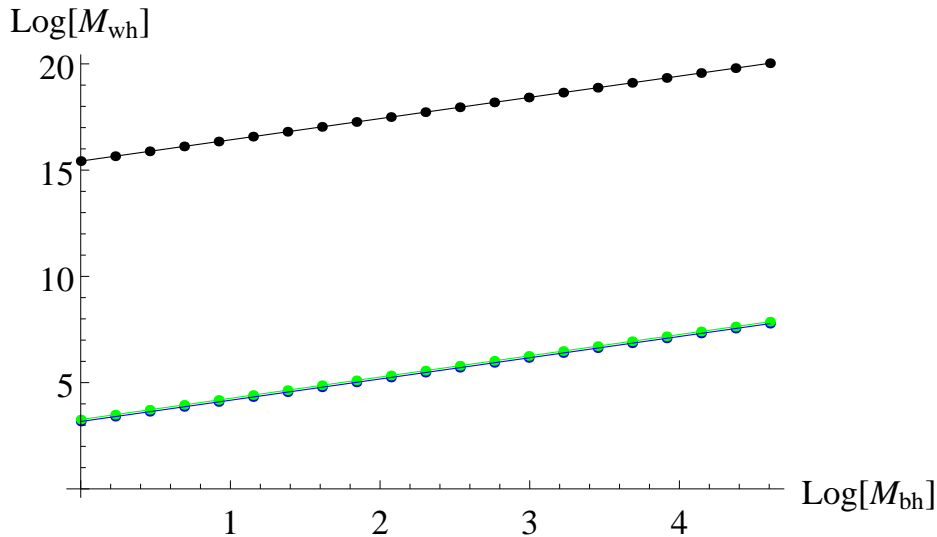}
	\end{centering}
	\caption{The logarithm of the mass associated to the ``white hole'' horizon $M_{\rm wh}$ is plotted versus the logarithm of the initial black hole mass $M_{\rm bh}$, ranging from $1$ to $100$ in Planck units. The black, blue and green dots correspond respectively to the masses obtained using the Hamiltonians $H_\text{eff}^{(1)}$, $H_\text{eff}^{(2)}$ and $H_\text{eff}^{(3)}$, with the fits given as $15.424 + 1.000\log[M_{\rm bh}]$, $3.171 + 1.000 \log[M_{\rm bh}]$ and $3.263 + 1.000\log[M_{\rm bh}]$ respectively. We take $G=c=1$, $\beta=0.2375$, $a^0=2$, $\mu_a=1/10$ and $\mu_b=1/10$.}
	\label{fig3}
\end{figure}

From the analysis of the evolution with different initial masses, the relations between black hole mass and the mass associated to the ``white hole'' horizon, induced by the different Hamiltonians we considered, have the same dependence on $M_{\rm bh}$, namely $M_{\rm wh}\propto M_{\rm bh}$ (see figure \ref{fig3}).
At this point, it is important to stress again that in general the relation between the initial black hole mass and the mass associated to the ``white hole'' horizon strongly depends on the choice of the initial condition $a^0$, which classically is an irrelevant quantity (due to the fact that, under coordinate transformation $R \to \lambda R$, the variable $a$ scales as $a \to \lambda^{-1} a$), and on the choice of the $\mu$ parameters. In particular, the relation $M_{\rm wh}\propto M_{\rm bh}$ is a consequence of the choice \eqref{InitCond} for $a^0$, and that the $\mu$ parameters are fixed numbers independent of the phase space
\footnote{Different values of the $\mu$ parameters or $a^0$ do not alter neither the qualitative properties of the evolution, nor the relation between the initial black hole mass and the mass associated to the ``white hole'' horizon.}.
But for instance, if one takes $\mu_b$ to be dependent of the black hole initial mass as $\mu_b\propto M_{\rm bh}^{-1}$, while keeping $\mu_a$ and $a^0$ as above, then in the case of the Hamiltonian $H_\text{eff}^{(1)}$ one obtains $M_{\rm wh}\propto M_{\rm bh}^5$. If one additionally takes $a^0$ to be proportional to $M_{\rm bh}^{-1}$, one obtains $M_{\rm wh}^{(1)} \propto M_{\rm bh}^4$. However, unlike the case shown in figure \ref{fig3}, the white hole mass dependence on $M_{\rm bh}$ in case of the Hamiltonians $H_\text{eff}^{(2)}$ and $H_\text{eff}^{(3)}$ will significantly differ from the one obtained with $H_\text{eff}^{(1)}$.

Lastly, we would like to make a general comment about such effective models: in standard quantum mechanics and field theory, the continuum limit is the only physical limit and the regulator is always taken to zero. In the context of loop quantum cosmology, the regulator is considered to be a physical fundamental scale, and therefore the limit of the regulator going to zero is considered to be unphysical. If this limit was taken, one simply recovers the classical general relativity Hamiltonian for all the effective Hamiltonians.
Unfortunately, there is so far no criterion to select the ``right" effective Hamiltonian. From the point of view of the theory, they are all eligible and on the same footing. The hope was that these different regularizations provide rather similar dynamics, but since this is not the case, this fact then brings to light an important ambiguity which must be studied and understood further, and other elements or procedures (such as renormalization in the full theory \cite{Bahr14,LLT1,BRS18}) need to be introduced in order to try to restrict it. This question however is beyond the scope of the current article.
\section{Summary \& discussion}\label{sec5}
In this paper we presented the construction of two new effective Hamiltonians for Kantowski-Sachs: $H_{\rm eff}^{(2)}$ is obtained following the prescription introduced in \cite{ADLPshort}, namely, regularizing the Euclidean part via \eqref{polymerization} and then using Thiemann identities for the Lorentzian part; $H_{\rm eff}^{(3)}$ is derived from the expectation value of Thiemann LQG Hamiltonian on a semiclassical state peaked on Kantowski-Sachs spacetime. We then compared the dynamics in the $\mu_o$-scheme generated by these effective Hamiltonians -- as well as the one generated by the effective Hamiltonian $H_{\rm eff}^{(1)}$ introduced in \cite{LQC-BH6} -- with initial conditions at the black hole horizon ($T = T_0 = 0$). However, since we are able to solve numerically the equations of motion of $H_{\rm eff}^{(3)}$ only to quadratic order in an expansion in the discreteness parameters, to make the comparison meaningful we have also considered a quadratic expansion of $H_{\rm eff}^{(1)}$ and $H_{\rm eff}^{(2)}$.
The analysis reveals that the integral curves of $H_{\rm eff}^{(2)}$ and $H_{\rm eff}^{(3)}$ are similar, but not identical, while the curves of expanded $H_{\rm eff}^{(2)}$ are in good agreement with those of $H_{\rm eff}^{(3)}$. {This agreement may suggest that $H_{\rm eff}^{(2)}$ and the full non expanded form of $H_{\rm eff}^{(3)}$ agree; but comparison of the second order terms in the expansions of $H_{\rm eff}^{(2)}$ with those in $H_{\rm eff}^{(3)}$ shows that they are different. We must therefore conclude that they are indeed two different effective Hamiltonians, but which generate similar dynamics in the $\mu_o$-scheme for the initial conditions we considered}. On the other hand, all the aforementioned curves are qualitatively different from the ones of $H_{\rm eff}^{(1)}$. Nevertheless, in all the cases considered, the singularity is replaced with a black hole to ``white hole'' transition, and the relations between the initial black hole mass $M_{\rm bh}$ and the final ``white hole'' mass $M_{\rm wh}$ have the same form, that is $M_{\rm wh} \propto M_{\rm bh}$.

As mentioned earlier, our analysis is realized in the $\mu_o$-scheme. This scheme is known to produce unphysical results in cosmology (such as a bounce at an energy density much lower than the Planck scale). However, $\mu_o$-scheme models are useful in order to extract the qualitative behavior of such effective models and grasp an understanding of the modifications with respect to the classical theory. One can of course develop the models in the $\bar \mu$-scheme using the Hamiltonians we considered above. In fact, recent proposals \cite{LQC-BH8, LQC-BH9} study models based on the effective Hamiltonian $H_{\rm eff}^{(1)}$ with $\mu$'s being phase space functions. The authors then reach different conclusions about the relations between the initial black hole mass and the final ``white hole'' mass. Implementing the $\bar \mu$-scheme with the Hamiltonians $H_{\rm eff}^{(2)}$ and $H_{\rm eff}^{(3)}$ will require a more elaborate analysis, and it is currently under investigation.

Finally, let us mention that our study was limited to spherically symmetric and static spacetime, namely Schwarzschild black hole interior. Further studies following a similar construction could be carried out in the context of non-static solutions, in which case one has to include the exterior of a black hole where the radial dependence cannot be ignored. While this is technically more involved, if successful, it would give us access to modeling physically realistic black holes, describing for example a spherical collapse. It might also give us a way to quantify the transition time involved in the black hole $\to$ "white hole" process, which would allow to make falsifiable predictions.
\begin{acknowledgments}
The authors would like to thank Jerzy Lewandowski for illuminating discussions. M.\ A.\ acknowledges the support of the project BA 4966/1-1 of the German Research Foundation (DFG) and the support of the Polish Narodowe Centrum Nauki (NCN) grant 2011/02/A/ST2/00300.  K.\ L.\ thanks the German  National  Merit  Foundation and NSF grant PHY-1454832 for their financial support.
\end{acknowledgments}

%
%

%
%
%
\end{document}